\documentclass[useAMS, usenatbib]{mn2e}
\usepackage{amsmath}
\usepackage{amssymb}
\usepackage[dvips]{graphicx}

\title[Internal shocks in jets]{iShocks: X-ray binary jets with an internal shocks model}
\author[O. Jamil , R. Fender and C. Kaiser]{O. Jamil$^{1}$ \thanks{E-mail:
oj1@soton.ac.uk}, R. P. Fender$^{1}$ and C. R. Kaiser 
\footnotemark[0]\\  
$^{1}$University of Southampton, U.K.\\
}

\begin{document}    
\date{Accepted 03 September 2009}

\pagerange{\pageref{firstpage}--\pageref{lastpage}} \pubyear{2009}

\maketitle

\label{firstpage}

\begin{abstract}
In the following paper we present an internal shocks model, iShocks,
for simulating a variety of relativistic jet scenarios; these
scenarios can range from a single ejection event to an almost
continuous jet, and are highly user configurable. Although the primary
focus in the following paper is black hole X-ray binary 
jets, the model is scale and source independent and could be used for
supermassive black holes in active galactic nuclei or other flows such
as jets from neutron stars. Discrete packets of plasma (or `shells')
are used to simulate the jet volume. A two-shell collision gives rise
to an internal shock, which acts as an electron re-energization
mechanism. Using a pseudo-random distribution of the shell properties,
the results show how for the first time it is possible to reproduce a
flat/inverted spectrum (associated with compact radio jets) in a
conical jet whilst taking the adiabatic energy losses into
account. Previous models have shown that electron re-acceleration is
essential in order to obtain a flat spectrum from an adiabatic conical
jet: multiple internal shocks prove to be efficient in providing this
re-energization. We also show how the high frequency turnover/break in
the spectrum is correlated with the jet power, $\nu_b \propto
L_{\textrm W}^{\sim 0.6}$, and the flat-spectrum synchrotron flux is
correlated with the total jet power, $F_{\nu}\propto L_{\textrm
  W}^{\sim 1.4}$ . Both the correlations are in agreement with
previous analytical predictions.
\end{abstract}


\section{Introduction}
In recent years there has been a lot of interest in
probing the disc-jet connection in a variety of astrophysical
objects. The
mechanisms behind the jet formation are still not fully understood, leaving many
open questions: the origin of a flat spectrum ($\alpha \sim
0$ when $F_\nu \propto \nu^{\alpha}$) is one such
question. The flat spectra have been observed both in
active galactic nuclei (AGN) (for a review see \citealp{cawthorne91}) and
X-ray binaries (XRBs) \citep{fender01}; in the XRBs
it has been seen to extend from Radio to
near infra-red \citep{corbel02}. It is thought that this spectrum
originates from the jet via the partially self absorbed synchrotron emission. \\
\indent The \citet{blandford79} model attempts to explain how such a flat
spectrum could arise. In their model, the jet is assumed to be conical with the
magnetic field perpendicular to the jet axis and frozen in plasma.
For a given frequency, an increase in the jet
volume causes a decrease in the plasma optical depth. The inner, denser, parts
of jets are optically
thick to lower frequencies (e.g. radio): the higher the energy density of
the jet volume, the higher the optical depth (for a given
frequency). The radio
frequencies therefore peak in the outer parts of the
jet, while the infra-red peak in the
inner parts of the jet. The
one drawback of the \citet{blandford79} model is the artificial replenishment of the
adiabatic energy losses suffered by the jet plasma. \citet{marscher80} showed
a model that takes electron energy losses into account, but are unable
to reproduce a flat spectrum; \citet{hjellming88}
presented a model where it is possible
to obtain a flat spectrum under slowed
expansion for a conical jet. A comprehensive study by \citet{kaiser06}
(more recently \citealp{pe'er09}) shows that if the adiabatic losses
are not replenished then it is impossible to obtain a flat spectrum
from a conical jet. \citet{kaiser06} also show that using a
non-conical jet volume, it is possible to recover a flat spectrum even
with energy losses: the jet geometry requires fine tuning to minimize
the adiabatic losses normally associated with a conical jet, while
allowing enough of a change in volume to drive the changes in optical depth. \\
\indent In the following paper, we present a model that addresses the
problem of electron
re-energization via a large number of `small' shell
collisions. Although the main focus in this paper is on black hole XRB
jets, the model is not restricted to such systems: the model is scale
independent and AGN jet volumes can also be simulated. The first part
of the paper goes through the details of the
model, outlining the physics and the techniques employed. The results
section is split into different jet scenarios: single ejection,
double ejection (single collision/internal shock), and multiple ejections (multiple
internal shocks). The results from these increasingly complex
scenarios are used to demonstrate the model's capabilities, in addition to
exploring the internal physics of the relativistic jets.   

\section{The Model}
\begin{figure*}
\includegraphics[width=0.75\textwidth]{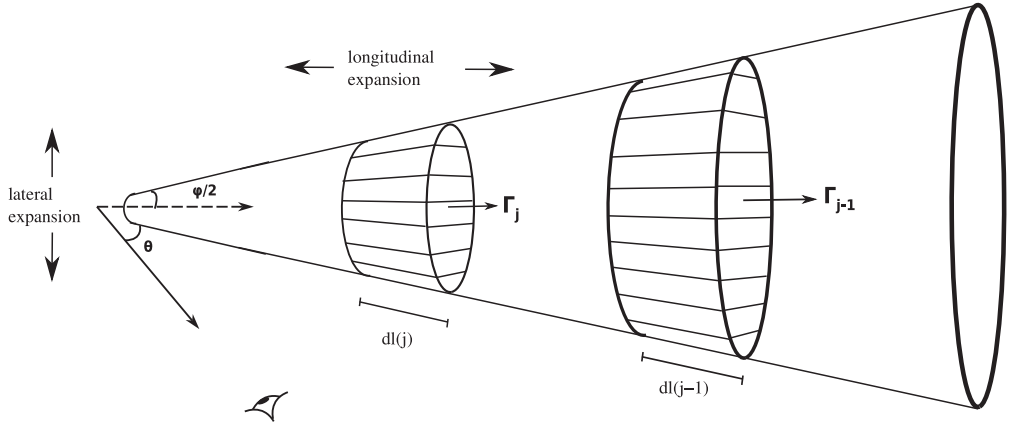}
\caption{An illustration of shells in our jet model. If the outer
  boundary of the inner shell, (j), contacts the inner boundary of the
  outer shell, (j-1), a collision is said to occur. The lateral
  expansion is due to jet opening angle; the longitudinal expansion is
  due to the shell walls expanding within the jet. The illustration
  is not to scale.}
\label{fig:schematic}
\end{figure*}

Our model is based on the \citet{spada01} internal shocks
model for radio-loud quasar. Many modifications, however, have
been carried out to make the model more flexible, and applicable to
different scales and scenarios. In our model the jet is
simulated using discrete packets of plasma or \textit{shells}. For
simplicity, only the
jets at relatively large angle of sight are treated. Each shell represents the
smallest emitting region and the resolution in the model is limited to
the shell size. While the simulation is running, the jet
can `grow' with the addition of shells at the base as the previously
added shells move further down the jet. If the time interval between
consecutive shell injections is kept small, a continuous-jet
approximation is achieved. The variations in shell injection time
gap and velocity cause faster shells to catch up with slower ones, leading
to collisions: the internal shocks, discussed later, are a result of
shell collisions. A schematic of the model set up is shown in figure
\ref{fig:schematic}: the two conical frusta shown represent the shells.

\subsection{Shell properties} 
The shell volume is based on a conical frustum (cone opening angle
= jet opening angle, $\varphi$). As a shell moves down the jet, it can
expand laterally as well as longitudinally (figure
\ref{fig:schematic}). The adiabatic energy losses are a result of the
work done by a shell in expanding; implicit
assumptions are made about the pressure gradient across the jet
boundary that would result in a conical jet. The emitting electron
distribution is assumed to be power-law in nature; each shell
contains its own distribution. The power-law distribution is of the form:
\begin{eqnarray}
  N(E) \textrm{d} E= \kappa E^{-p}\textrm{d}E \ ,
\label{eqn:plaw}
\end{eqnarray}
where $E=\gamma m c^2$ is the electron energy, $p$ is the power-law
index and $\kappa$ is the normalization factor. If the total kinetic
energy density of
the electrons, $E_k$, is known then $\kappa$ can be calculated for the
two cases of power-law index: $p\neq2$, and $p=2$. When $p\neq2$, we
have (with the electron energy is expressed in terms of the Lorentz
factor with $mc^2=1$):
\begin{eqnarray}
  E_k
  &=&\kappa\left[\frac{1}{(2-p)}(\gamma_{max}^{(2-p)}-\gamma_{min}^{(2-p)})\right .
    \nonumber \\
   &-& \left .\frac{1}{(1-p)}(\gamma_{max}^{(1-p)} -
    \gamma_{min}^{(1-p)})\right] \ ,
\label{eqn:pnorm1}
\end{eqnarray}
and for $p=2$:
\begin{eqnarray}
E_k=\kappa\left\{[\textrm{ln}(\gamma_{max}) -
    \textrm{ln}(\gamma_{min})]
    + [\gamma_{max}^{-1}-\gamma_{min}^{-1}]\right\} \ ,
\label{eqn:pnorm2}
\end{eqnarray}
where the subscripts $max$ and $min$ denote the upper and lower energy
bounds for the electron distribution. The relations given in equations
\ref{eqn:pnorm1} and \ref{eqn:pnorm2} can therefore be used to
calculate the change in
electron power-law distribution when there is a change in the total
kinetic energy density, assuming the power-law index and $\gamma_{min}$ are
fixed. $\gamma_{min}$ value throughout the following work is
  set equal to unity, while the power-law index is assumed to be
  2.1. The electron energy distribution upper limit, $\gamma_{max}$, is
  initially set to be $10^6$, but allowed to
  vary with the energy losses.

A magnetic field is essential to give rise to the synchrotron
radiation. In the shells, the magnetic field is assumed to be
constantly tangled in the plasma, leading to an assumption
that the magnetic field is isotropic; hence, treated like an
ultra-relativistic gas \citep{heinz00}. If the magnetic energy
density ($E_B$) is given, the field ($B$) can be calculated:
\begin{eqnarray}
E_B=\frac{B^2}{2\mu_0} \ ,
\label{eqn:magdens}
\end{eqnarray}
where $\mu_0$ is the magnetic permeability.

Other shell properties include the bulk Lorentz
factor, $\Gamma$, and the shell mass, $M$. If there is a variation in
the $\Gamma$ of
different shells in the jet, then the faster inner shells are able to
catch up with the slower outer ones, causing shell collisions; the
shell collisions create internal shocks, which ultimately generate the internal
energy.

\subsection{Internal shocks}
\begin{centering}
\begin{figure}
\includegraphics[width=0.45\textwidth, totalheight=0.48\textheight]{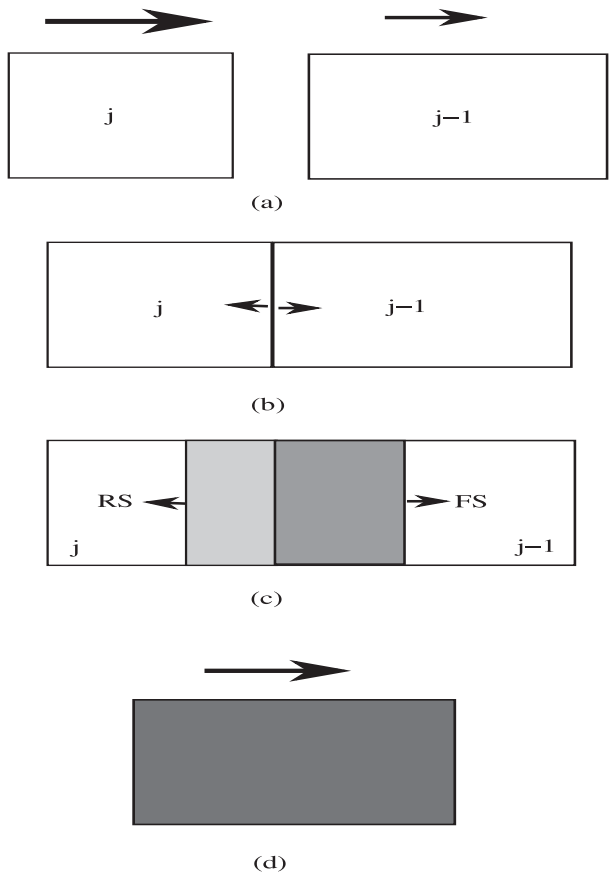}
\caption{An illustration of a two shell collision leading to a
  merger: (a) when the inner shell, j, comes in contact with the
  outer shell, j-1, (b) a shock starts to form; (c) the forward shock, FS,
  travels through the outer shell while the reverse shock, RS, travels
through the inner shell; (d) once shock fronts have traversed the two
shells, a new `merged' shell is formed.}
\label{fig:shellColl}
\end{figure}
\end{centering}

When two shells collide, a shock forms at the contact surface. Some of
the steps involved in two-shell collision, and the subsequent merger, are
shown in figure \ref{fig:shellColl}. The
collision are considered to be inelastic. With many shells present inside
the jet, first we need to calculate the next collision time between two
shells: a collision is said to occur when the outer boundary of the
inner shell, $R_{j}^{outer}$, comes in contact with the inner
boundary of the outer shell, $R_{j-1}^{inner}$. The following relation
can be
used to calculate the time interval for two shell collision:
\begin{eqnarray}
  \textrm{d}t_{coll}=\frac{R_{(j-1)}^{inner} -
    R_{(j)}^{outer}}{(\beta^e_{(j-1)}+\beta^e_{(j)})c +
    (\beta_{(j)}-\beta_{(j-1)})c} \ ,
\label{eqn:colltime}
\end{eqnarray}
where the subscripts $j-1, j$ denote two consecutive shells, $\beta_e$
is the shell longitudinal expansion velocity (along the jet axis), and 
$\beta$ is the shell velocity
($=\sqrt{\Gamma^2-1}/\Gamma$); $\textrm{d}t$ is calculated for all the
shells inside the jet and the minimum of these collision time gaps is
selected i.e. the next earliest collision. The
shell longitudinal expansion ($\beta_e$) is due to any thermal energy the shell may
have. We do not explicitly model a thermal electron population;
however, the expansion effects of having such a population are
incorporated in the model. The shell expansion velocity is given by \citep{spada01}:
\begin{eqnarray}
  \beta^e = \frac{2\beta_s'}{\Gamma^2}\frac{1}{1-(\beta\beta_s')^2} \ ,
\label{eqn:betaexp}
\end{eqnarray}
where $\beta_s=v_s'/c$ corresponds to the sound velocity in the plasma
(in shell co-moving frame), and:
\begin{eqnarray}
  v_s'=\sqrt{\frac{1}{3}\frac{E_{th}'}{M}} \ ,
\label{eqn:soundv}
\end{eqnarray}
with $E_{th}'$ being the shell thermal energy and M is the shell
mass. The prime denotes quantities in the shell rest frame.
The \citet{panaitescu99} treatment of
shock propagation is followed to work out the various quantities
associated with the shock itself. The shock propagation can be split
into two shock-fronts. The forward shock travelling from the contact surface
and through the outer shell $(j-1)$. The reverse shock travels through
the inner shell $(j)$ (injected after shell $(j-1)$). Once the
shock-front has passed through, the plasma/shell is considered to be
\textit{shocked} (S) and will have different physical properties
compared to the unshocked
plasma (see figure \ref{fig:shellColl}). In one of the shell
co-moving frames (shell rest-frame), the shock-front (SF) velocity can be calculated as:
\begin{eqnarray}
  \beta_{SF}'=\frac{(\Gamma_S'-1)(\hat{\gamma}\Gamma_S'+1)}{\beta'\Gamma_S'
    [\hat{\gamma}(\Gamma_S'-1)+1]} \ ,
\label{eqn:betashock}
\end{eqnarray}
$\hat{\gamma}$ is the adiabatic index and $\Gamma_S'$ corresponds to
the shocked plasma and is given by:
\begin{eqnarray}
  \Gamma_S'=\Gamma_m\Gamma(1-\beta_m\beta) \ ,
\label{eqn:shockedG}
\end{eqnarray}
where the subscript $m$ denotes the merged shell properties. The merged shell is
formed once the shock-fronts have passed through both shells, leaving
one combined shell. The merged shell mass is simply the linear
combination of the two merging shells i.e.
\begin{eqnarray}
  M_m=M_{(j)}+M_{(j-1)} \ ,
\label{eqn:mergemass}
\end{eqnarray}
while the merged shell Lorentz factor is given by \citep{spada01}:
\begin{eqnarray}
  \Gamma_m=\left(\frac{\mu_{(j)}\Gamma_{(j)}+\mu_{(j-1)}\Gamma_{(j-1)}}
        {\mu_{(j)}/\Gamma_{(j)}+\mu_{(j-1)}/\Gamma_{(j-1)}} \right)^{1/2}
        \ ,
\label{eqn:merglorentz}
\end{eqnarray}
where $\mu = M + \eta/c^2$ and $\eta$ is the shell internal energy. As
the shell collisions are considered inelastic, the
generated internal energy is given by:
\begin{eqnarray}
  E_{in} &=&
  \eta_{(j)}+\eta_{(j-1)}+\mu_{(j)}c^2(\Gamma_{(j)}-\Gamma_{m})
  \\ \nonumber &+& \mu_{(j-1)}c^2 (\Gamma_{(j-1)}-\Gamma_{m}) \ .
\label{eqn:merintene}
\end{eqnarray}

Once the quantities, outlined above, associated with the shocked plasma are
calculated, we are able to determine the new merged shell length. Due to
the shock propagation through the plasma, the two shell lengths cannot
be combined linearly to give the merged shell length. The shock
propagation has a compression effect on the shell; the
merged shell length is given by:
\begin{eqnarray}
  \textrm{d}l_m = \frac{\textrm{d}l_{(j)}}{\rho_{(j)}} +
  \frac{\textrm{d}l_{(j-1)}}{\rho_{(j-1)}} \ ,
\label{eqn:mergwidth}
\end{eqnarray}
and the density, $\rho$, is:
\begin{eqnarray}
  \rho =
  \frac{\Gamma_m}{\Gamma}\frac{\hat{\gamma}\Gamma_S'+1}{\hat{\gamma}-1} \ .
\label{eqn:mergrho}
\end{eqnarray}
The Lorentz factor, $\Gamma$, corresponds to one of the two shells involved in
the collision while $\Gamma_S'$ is given by equation
\ref{eqn:shockedG}. We do not consider the re-energization of the new
merged shell to be instantaneous, but instead the energy is dissipated
over a time period the shock-fronts would take to cross the inner and outer
shells combined i.e. 
\begin{eqnarray}
  \textrm{d}t_{ER}=\frac{\textrm{d}l_{(j)}}{\beta_{RS}}+\frac{\textrm{d}l_{(j-1)}}{\beta_{FS}}
      \ ,
\label{eqn:dtER} 
\end{eqnarray}
where the subscripts $ER, FS$ and $RS$ denote energy release
(time), forward shock and reverse shock respectively. Due to the
limitations in the way the shocks are modelled, the merged shell is
given the corresponding length (equation \ref{eqn:mergwidth}) at the point
of creation; only the energy release is delayed over
time d$t_{ER}$. If the adiabatic energy losses are taken into account,
then the merged shell will also be losing energy during the period d$t_{ER}$. 

\subsection{Adiabatic losses}
For an expanding conical jet, the adiabatic energy
losses need to be taken into account. These energy losses are due to the work done by
the jet material while expanding
($\textrm{d}U=-P\textrm{d}V$, where U, P and V are energy, pressure
and volume respectively). Using the relation, $P=nkT$, and assuming
that all the synchrotron emitting electrons are highly relativistic, we have:
\begin{eqnarray}
 \frac{\gamma}{\gamma_0}=\left(\frac{V}{V_0}\right)^{-\frac{1}{\hat{\gamma}}} \ ,
\label{eqn:adiaeleLorentz}
\end{eqnarray}
with the adiabatic index, $\hat{\gamma}=3$; the subscripts '0' denote
quantities before the change in volume; the
Lorentz factor $\gamma$ corresponds to the power-law electrons
accelerated to high energies due to the shock-front passing through the
plasma. The adiabatic energy losses for the kinetic energy contained in the
power-law electron distribution can therefore be calculated using equation
\ref{eqn:adiaeleLorentz}. To calculate the change in the power-law normalization
, $\kappa$, for a change in volume, the following
relation is used:
\begin{eqnarray}
  \kappa = \kappa_0\left(\frac{V}{V_0}\right)^{\frac{-p-2}{3}} \ .
\end{eqnarray}
Once the power-law normalization is, initially, calculated using
equation \ref{eqn:pnorm1} or \ref{eqn:pnorm2}, the above relation can then
be used to calculate the subsequent changes in the normalization. If the shell is
involved in another collision then the distribution is re-calculated completely.
The change in
the maximum Lorentz factor of the electrons,
$\gamma_{max}$, as the shell volume changes, can be calculated using equation
\ref{eqn:adiaeleLorentz}. The combined effect
of varying $\gamma_{max}$ and $\kappa$ is to effectively `evolve' the
power-law electron distribution.

The changes in the magnetic energy density can be
determined in a similar manner. If we assume that the magnetic field
is constantly entangled in the plasma
and treat it as an ultra-relativistic gas, we can calculate the
changes in magnetic pressure, $P(B)$ using:
\begin{eqnarray}
PV^{\hat{\gamma}}=P_0V_0^{\hat{\gamma}}  \ ,
\label{eqn:pv}
\end{eqnarray}
therefore,
\begin{eqnarray}
P(B)=P_0(B)\left(\frac{V}{V_0}\right)^{-\frac{1}{\hat{\gamma}}} \ ,  
\label{eqn:magp}
\end{eqnarray}
where the adiabatic index $\hat{\gamma} = 4/3$.

\subsection{Partially self absorbed synchrotron emission}
To model the synchrotron radiation, we employ the treatment outlined in
\citet{longair94}. With only the power-law electrons present, the
synchrotron emission calculation is simplified; the synchrotron
monochromatic intensity is given by:
\begin{eqnarray}
  I_{\nu}=\delta_{\mp}^3\frac{J_{\nu}}{4\pi\chi_{\nu}}(1-e^{-\chi_{\nu}r})
  \ .
\label{eqn:synchrI}
\end{eqnarray}
The emission coefficient, $J_{\nu}$, and the absorption coefficient,
$\chi_{\nu}$, are given by \citet{longair94}. These coefficients
are a function of the power-law normalization and the magnetic field,
which in turn depends on the energy density of the shells; $r$ is the shell radius
($\tau_{\nu}=\chi_{\nu}r$); the Doppler factor,
$\delta_{\mp}$, is:
\begin{eqnarray}
  \delta_{\mp}=\left[\Gamma(1\mp\beta(\textrm{cos}\theta)\right]^{-1} \ ,
\label{eqn:doppler}
\end{eqnarray}
where, $\theta$ is the jet viewing angle and `$\mp$' corresponds
to either an approaching component or a receding
component of the jet.

If the shell area is given by $A$ and the distance to the jet
is $D$, then the flux, $\textrm{d}F_{\nu}$, from a single shell has
the following form:
\begin{eqnarray}
  \textrm{d}F_{\nu}=\delta_{\mp}^3\frac{A}{4\pi^2D^2}\frac{J_{\nu}}{\chi_{\nu}}(1-e^{-\chi_{\nu}r})
  \ .
\label{eqn:flux}
\end{eqnarray}
The above relations are used to self consistently
calculate the synchrotron spectrum as it varies with the shell
properties. A more indepth treatment of the radiative transfer
should take the relativistic effects, such as the relativistic
aberration, into account as they may affect the overall spectral
normalization. However, with simplicity in mind, these effects
are not taken into account here: any errors due to this approximation
can be minimized by only treating the jets at a large angle of
sight. 

When a shell expands, its optical depth (with
respect to a given frequency) changes; hence, a shell that is optically
thick to radio frequencies can become optically thin to them, as it
moves down the jet. The optical depth of the
neighbouring shells is not taken into account when calculating the
emission from a given shell; only the jets at relatively large
angle sight from the viewer can be modelled. In other words, only
the emission directly from each individual shell is modelled.

\subsection{Model parameters}
\label{subsec:modelparams}
So far, only some of the physics and the principles behind the model
have been outlined. It is therefore important to list some of the
parameters used in our model to see how they influence the physical
properties of a jet (appendix \ref{app:params} contains all the
model parameters). 

The internal energy of the shell is split
between the electron kinetic energy ($u_e$), shell thermal energy ($u_{th}$)and magnetic
energy($u_B$) i.e.
\begin{eqnarray}
  E_{int} = u_e + u_{th} + u_B \ .
\label{eqn:esplit}
\end{eqnarray}
Equipartition between the electron kinetic and magnetic energy is
assumed. The thermal energy density, $u_{th}$, is assumed to
  be solely responsible for the longitudinal expansion of the shells. Although the
power-law electrons would also exert pressure for a similar effect,
this becomes more important when $u_{th}$ is zero. However, any
\textit{realistic} jet scenario is modelled with the thermal
population present; the zero-thermal-energy scenario is used mainly
for demonstrative purposes. Also if the shell mergers are not taking
place and all the injected shells are identical, the longitudinal expansion
needs to be suppressed, otherwise
equation \ref{eqn:merglorentz} generates an
erroneous value for the merged $\Gamma_m$. Of course, a more detailed
treatment for any
future work, take into account not only the power-law
electron pressure, but also model the synchrotron
emission from the thermal population. 

It maybe
possible to constrain the distance to the
source ($D$) and  the jet viewing angle ($\theta$) from
observations. The shell mass on the other hand offers a free
parameter. In case of massive ejection event, the mass can be set
manually. For a continuous jet (multiple ejections), we have a choice
of setting the individual shells' properties manually or sampling from a
\textit{pseudo}-random distribution of these parameters. In the
\textit{pseudo}-random case, if the jet
kinetic luminosity, $L_W$, is known then it can be
used to generate the shell mass values by using:
\begin{eqnarray}
  \sum_{j=1}^N M_j \Gamma_j c^2 = L_Wt_{jet} \ ,
\label{eqn:jetkinetic}
\end{eqnarray}
where $t_{jet}$ is the duration the jet is ``on''. Therefore the total
relativistic mass of all the shells present in the jet has to
correspond to the jet kinetic luminosity
of the jet being modelled. The time gap,$d_{inj}$, between any two shells can
also be set either manually or sampled from a Gaussian distribution with a
given mean and standard deviation. The shell bulk Lorentz factor
($\Gamma$), if not set manually, is picked from a random distribution of
values with the maximum and the minimum values ($\Gamma_{max},
\Gamma_{min}$) set by the user; the
bulk Lorentz factor for a shell does not vary as it moves through
the jet, unless it is
involved in a collision. The shell
length ($\textrm{d}l$) and the jet opening angle ($\varphi$) are not
constrainable from observations. However, at least in the case of the shell
length, attempting to achieve a continuous jet approximation results
in the following relation:
\begin{eqnarray}
  dl = l_{scale} \ dt_{inj} \ \beta_{shell} c  \ ,
\label{eqn:shellLength}
\end{eqnarray}
where $l_{scale}$ is a scaling factor with a maximum of unity.
$l_{scale} =1$ means no spatial gap between two consecutively
injected shells; in the simulation, however, $l_{scale} < 1$ in order to avoid
a `pile-up' of shells at the source.

\section{Results}
With the aim of demonstrating a few of our model's capabilities, the
following section is split into three main scenarios:
single ejection, double ejection, and multiple ejections. Within each
scenario, the effects of adiabatic energy losses are also
explored. As stated earlier, the focus in the following sections is on
black hole XRB jets; accordingly various shell properties, outlined
above, are calculated and estimated for XRB scaled jets. 

\subsection{Single ejection}
A single shell ejection scenario can be used to simulate the massive ejection
events observed in X-ray binaries, in the form of radio flares, when
the source is transitioning
from a hard state to a soft state \citep{fender04}. When
shells are injected into the jet, they only start emitting
radiation, or ``light-up'', when they have been involved in a
collision i.e. the shock front has passed through and energized
the plasma within the shell (unless if they are injected with internal
energy). In the case of a single shell, however,
no collisions can take place; thus, the shell will require
lighting-up artificially. This is achieved by creating a
``shock-zone'' at an arbitrary point, $x_{shock}$, along the jet; the
\textit{shock-zone} picture is reminiscent of the jet models
involving a single-shock-zone in the jet (see \citealp{falcke96},
\citealp{falcke00}, \citealp{markoff01}, \citealp{markoff03},
\citealp{pe'er09} and the references therein). In our model, once a
shell passes through the shock zone it is energized
instantly. A fraction of the shell's relativistic energy, $E_{frac}$, is used as a
scaling for the amount of internal energy given to a shell after
energization; the shell velocity, $\Gamma$, and mass, $M$, remain unchanged.

\subsubsection{Without energy losses}
Not taking the shell energy losses, as it expands, into account is
similar to assuming that any energy losses are continually
replenished \citep{blandford79}. With this kind of set up, a shell
is allowed to propagate down the jet and expand laterally. The shell longitudinal
expansion, due to the thermal energy of the plasma, is suppressed: the
thermal energy is set to zero. The change in volume is therefore not
associated with any work done by the shell.
\begin{figure}
\includegraphics[width=0.48\textwidth]{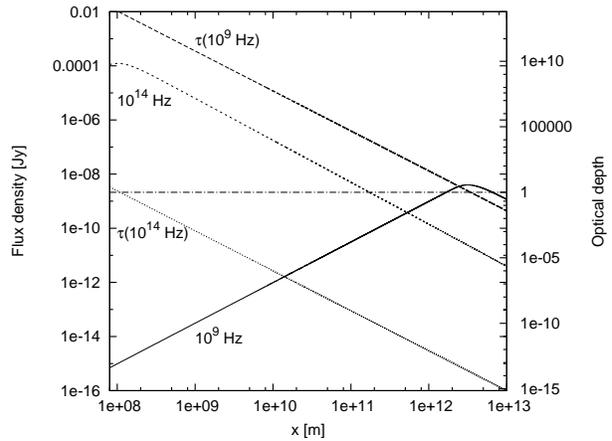}
\caption{Emission from a single shell: radio and infra-red frequencies
  are shown. The shell optical depth, $\tau_{\nu}$, is also shown; the
  horizontal line marks optical depth of unity.}
\label{fig:singleNoloss}
\end{figure}

\begin{figure}
\includegraphics[width=0.47\textwidth]{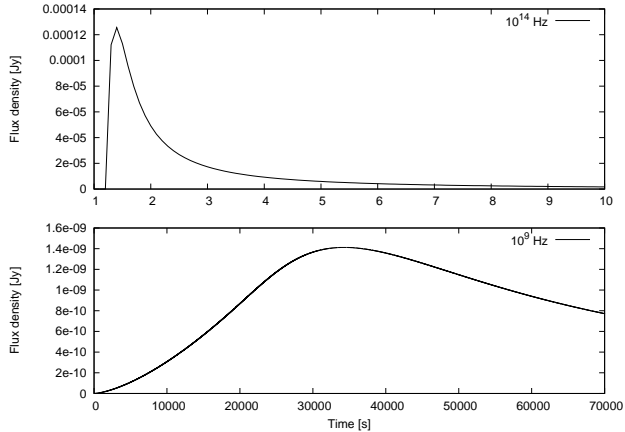}
\caption{Infra-red and radio lightcurves for a single shell without
  adiabatic energy losses.}
\label{fig:singleLc}
\end{figure}

The emission at radio and infra-red frequencies, from a single shell, is shown in figure
\ref{fig:singleNoloss}. The plot also shows how the shell optical depth for
the two frequencies changes as the shell moves down the jet. The
increase in shell volume causes the shell to become optically thinner
to lower frequencies. This behaviour is evident in the figure, as the radio
emission peak is much further down the jet than the infra-red peak:
the emission peaks at $\tau_{\nu}\approx 1$. Various shell and
jet volume parameters are outlined in table \ref{tab:figsparamsingle}.
\begin{table}
\caption{The parameters used for single ejection scenarios.}
\begin{center}
\begin{tabular}{@{\extracolsep{\fill}}lcr@{}}
\hline
Parameter & fig \ref{fig:singleNoloss}, \ref{fig:singleLc}  & fig
\ref{fig:singleLoss}, \ref{fig:singleLossLc} \\
\hline
$\varphi$ & 5$^o$ & 5$^o$\\
$\theta$  & 40$^o$ & 40$^o$\\
$D$ & $2$ kpc & $2$ kpc \\
$M$ & $1\times 10^7$ kg & $1\times 10^7$ kg \\
$\Gamma$ & 2.0 & 2.0\\
$\textrm{d}l$ & $1\times10^4$ m & $1\times10^4$ m\\
$u_e$ & 0.5 & 0.5 \\
$u_B$ & 0.5 & 0.5 \\
$u_{th}$ & 0.0 & 0.0 \\
$x_{shock}$ & 0.2 ls & 0.2 ls\\
$E_{frac}$ & 0.3 & 0.3 \\
Sim. Duration & $7\times10^4$ s& $500$ s\\
\hline
\end{tabular}
\end{center}
\label{tab:figsparamsingle}
\end{table}

The radio and the infra-red lightcurves, in figure \ref{fig:singleLc},
illustrate how the radio rise time is much longer than that of
the infra-red. These rise times are determined by the shell energy
density, and how quickly this energy density changes with the change
in volume. Therefore, in the case shown here, the shell properties are
such that the infra-red peaks rapidly after the injection, but
the radio takes much longer. It should be noted that because energy losses
are not considered here, only the energy
density of the shell, due to lateral expansion, is decreasing; the
shell's total energy content remains fixed. 

\subsubsection{With adiabatic energy losses}
When the adiabatic energy losses are taken into account, an increase in
shell volume causes a decrease in the shell internal energy. In
this particular simulation, as in the previous scenario, the shell thermal energy is
again set to zero. Therefore any increase in the shell volume is
purely due to the lateral expansion of the jet.  
\begin{figure}
\includegraphics[width=0.48\textwidth]{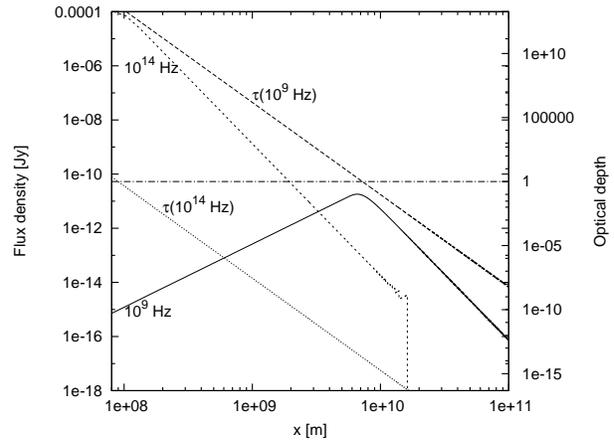}
\caption{Emission from a single shell with adiabatic energy losses:
  radio and infra-red frequencies
  are shown. The shell optical depth, $\tau_{\nu}$, is also shown; the
  horizontal line marks optical depth of unity.}
\label{fig:singleLoss}
\end{figure}

\begin{figure}
\includegraphics[width=0.47\textwidth]{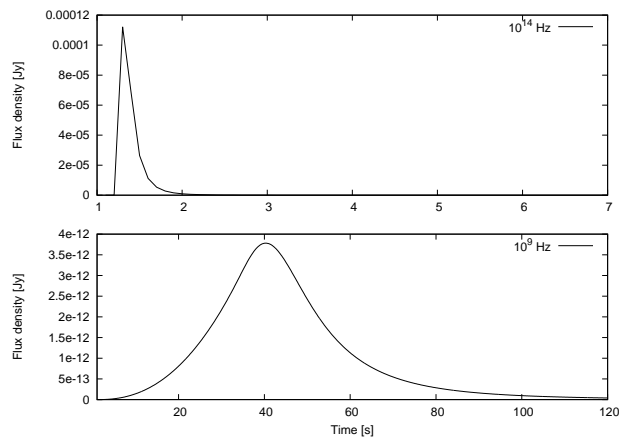}
\caption{Infra-red and radio lightcurves for a single shell with
  adiabatic energy losses.}
\label{fig:singleLossLc}
\end{figure}

The radio and infra-red spectra from a single shell, with the adiabatic
energy losses taken into account, are shown in figure
\ref{fig:singleLoss}. When compared with the spectra with no energy
losses, figure \ref{fig:singleNoloss}, there are two main differences
that become
apparent: the radio peak is at
much lower flux value, and at much smaller distance
along the jet. The infra-red peak flux values, however, remain relatively
unchanged; the peak occurs at smaller radii when compared to the no energy
losses case. These differences can be explained when one looks at how
the shell energy density changes with and without the energy losses present.    

The shell optical depth is a function of the shell's energy density,
i.e. a shell becomes optically thin to lower frequencies as the energy
density drops. When energy losses are not considered, the
change in the shell volume solely effects the change in the
energy density. However, when the adiabatic energy losses are taken
into account, the energy losses along with an increase in the shell
volume drive the change in the shell energy density. In such a
scenario, there is a two fold effect on the shell energy density:
increase in the shell volume and the decrease in the shell internal
energy. This then means that the shell optical depth, for a given
frequency, changes more
rapidly than when only the volume is effecting the change. The
emission intensity, however,
is also affected by the energy losses. With the adiabatic energy
losses active a shell may be able to  peak, for example, in radio
frequencies at smaller jet radii, $x$, but the peak
intensity is lower due to the energy losses suffered by the
emitting electrons and the magnetic field.(If longitudinal expansion
is also taken into account then the energy losses are accelerated and
a shell is able to peak in radio frequencies even earlier with a
further decrease in flux values). This is can be seen in
figure \ref{fig:singleLoss}, and the lightcurves in figure
\ref{fig:singleLossLc}, where the peak radio flux is nearly two
orders of magnitude lower when compared with the peak radio flux in figure
\ref{fig:singleNoloss}. The infra-red peak flux on the other hand suffers a
relatively a small reduction. This is due to the shell becoming optically thin to IR
frequency very quickly in both cases, thus not having the time to
suffer much energy energy losses. When the adiabatic losses are
active, the shell starts off optically thin to IR. This is
because the relative change in volume from the moment of injection to the
subsequent time step being sufficient to drop the shell energy density
below the limit for the shell becoming optically thin to infra-red
frequencies. The initial volume of a shell, in addition to relative
change in volume, also
plays a part in how quickly that shell becomes optically thin
to a given frequency: a large enough shell could start off as
optically thin to infra-red frequencies. The sharp cut off for the
infra-red flux, in figure 4, is due to $\gamma_{max}$ dropping below
the energy threshold for the power-law electrons to emit in
the infra-red. 

\subsection{Double ejection}
A double ejection scenario involves two shells being injected into the
jet volume with a time interval, $\textrm{d}t_{inj}$, between them. If the second shell,
$j$, has a higher velocity than the preceding shell, $j-1$, then the
two shells should eventually collide. This scenario also demonstrates the core
principle of multiple ejections: a large number of
two-shell collisions taking place all along the jet to give rise to
multiple shocks.

The shock-zone location, mentioned in the previous scenario, is set at
zero. This means that the shells are injected with some internal
energy instead of gaining it after passing through an arbitrary
point. The adiabatic losses are also modelled, but only due to lateral
expansion of the jet; the shell thermal energy is set to be zero.  

\begin{figure}
\includegraphics[width=0.48\textwidth]{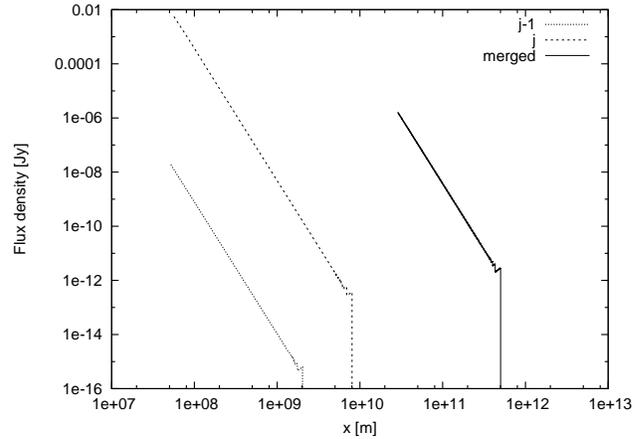}
\caption{The infra-red ($1\times10^{14}$ Hz) emission from the two
  injected shells that later on
merge (at $\sim$100s) to become a single shell. The properties of the two injected shells are
outlined in table \ref{tab:twoshells} while the simulation parameters
are outlined in table \ref{tab:doubleejection}. Adiabatic energy losses due to lateral
expansion only are being modelled.} 
\label{fig:doubleloss_ir}. 
\end{figure}

\begin{figure}
\includegraphics[width=0.48\textwidth]{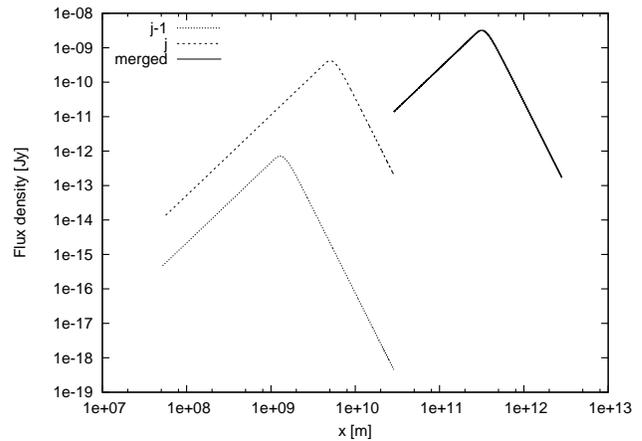}
\caption{The radio ($1\times10^{9}$ Hz) emission from the two shells injected that later on
merge (at $\sim$100s) to become a single shell. The properties of the two shells are
outlined in table \ref{tab:twoshells} while the simulation parameters
are outlined in table \ref{tab:doubleejection}. Adiabatic energy losses due to lateral
expansion only are being modelled.}
\label{fig:doubleloss_rad}
\end{figure}

\begin{figure}
\includegraphics[width=0.48\textwidth]{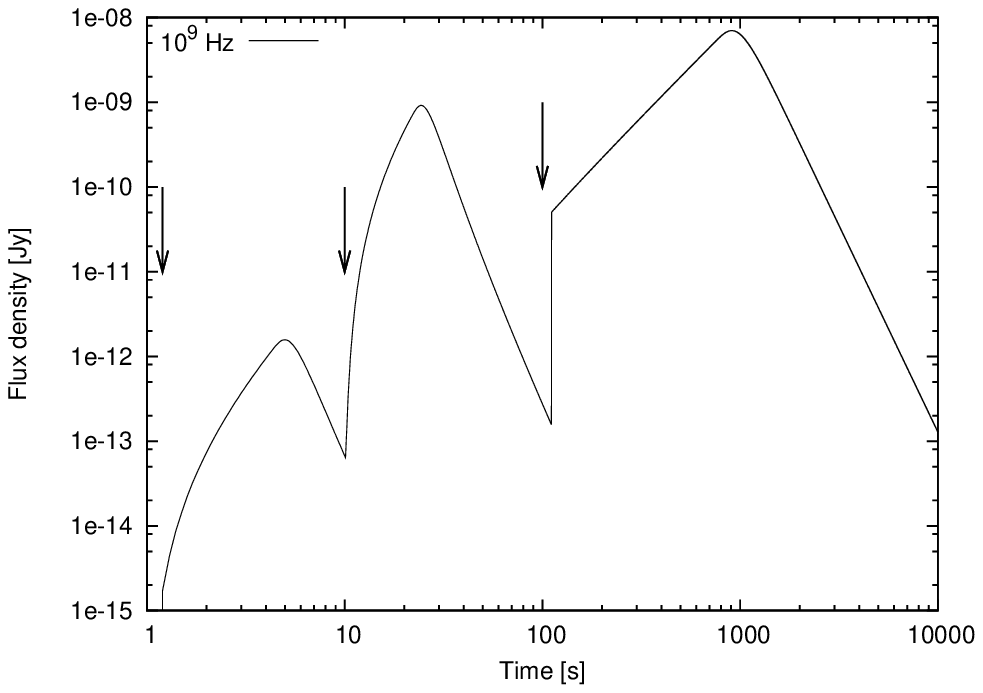}
\caption{The lightcurve for the radio
  emission. The two arrows on the left signify injection of the two
  shells; the third arrow shows the time of merger. Initially (after
  the second shell injection at 10s) the
  lightcurve comprises emission from both the shells; later, after
  merger at $\sim$ 100s, only a
  single shell exists in the jet. The properties of the two shells are
outlined in table \ref{tab:twoshells}; the simulation parameters
are outlined in table \ref{tab:doubleejection}. The unusual
\textit{log}(time) is used for demonstrative purposes.}
\label{fig:doubleloss_lc_rad}
\end{figure}

\begin{figure}
\includegraphics[width=0.48\textwidth]{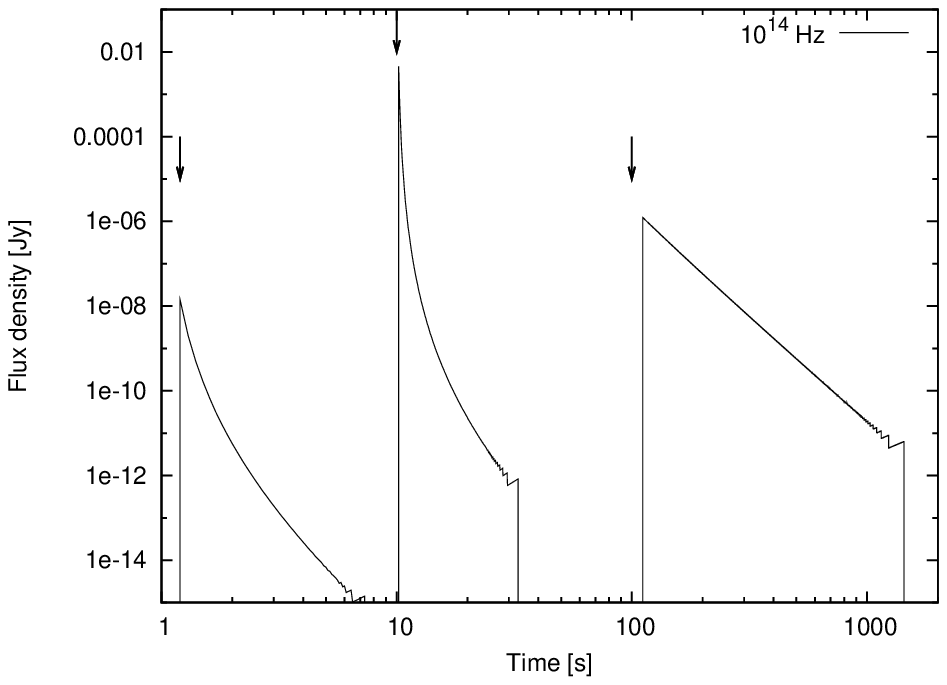}
\caption{The lightcurve for the infra-red
  emission. The two arrows on the left signify injection of the two
  shells; the third arrow shows the time of merger. Initially (after
  the second shell injection at 10s) the
  lightcurve comprises emission from both the shells; later, after
  merger at $\sim$ 100s, only a
  single shell exists in the jet. The properties of the two shells are
outlined in table \ref{tab:twoshells}; the simulation parameters
are outlined in table \ref{tab:doubleejection}. The unusual
\textit{log}(time) is used for demonstrative purposes.}
\label{fig:doubleloss_lc_ir}
\end{figure}

\begin{table}
\caption{The properties of the two shells injected in the
  double-ejection scenario shown in figures \ref{fig:doubleloss_ir},
  \ref{fig:doubleloss_rad}, \ref{fig:doubleloss_lc_rad}and
  \ref{fig:doubleloss_lc_ir}}
\begin{center}
\begin{tabular}{@{\extracolsep{\fill}}lcc@{}}
\hline
Parameter & shell 1 & shell 2 \\
\hline
Inj. time & 1 s & 10 s\\
Mass  & $1\times 10^{6}$ kg & $1\times 10^{10}$ kg \\
$\Gamma$ & 2.0 & 3.0\\
$\textrm{d}l$ & $1\times 10^{5}$ m & $1\times 10^{7}$ m\\
\hline
\end{tabular}
\end{center}
\label{tab:twoshells}
\end{table}

\begin{table}
\caption{The parameters used for double-ejection scenario.}
\begin{center}
\begin{tabular}{@{\extracolsep{\fill}}lr@{}}
\hline
Parameter & Figures \ref{fig:doubleloss_ir}, \ref{fig:doubleloss_rad},
\ref{fig:doubleloss_lc_rad} and \ref{fig:doubleloss_lc_ir}\\
\hline
$\varphi$ & 5$^o$ \\
$\theta$  & 40$^o$ \\
$D$ & $2$ kpc \\
$u_e$ & 0.5 \\
$u_B$ & 0.5 \\
$u_{th}$ & 0.0 \\
$x_{shock}$ & 0.1 ls \\
$E_{frac}$ & 0.2 \\
Sim. Duration & $1\times10^4$ s \\
\hline
\end{tabular}
\end{center}
\label{tab:doubleejection}
\end{table}

The properties of the two shells injected into the jet volume are
outlined in table \ref{tab:twoshells}. The first shell, $(j-1)$, is not only
less massive than the following one, $j$, but also larger. The combination
of these parameters means that $(j-1)$ becomes optically thin to lower
frequencies sooner than $(j)$. This can be seen in the
figures \ref{fig:doubleloss_ir} and  \ref{fig:doubleloss_rad} where
the radio peak for shell $(j-1)$ is at much smaller jet radii than that
for shell $j$; in the case of IR emission, shell $(j-1)$ is already optically
thin at those frequencies, when it is injected, while $j$ reaches the
peak IR flux later. The figures also show the point where the
two-shell collision, or merger, takes place: it is marked by a sharp
increase in both the radio and the IR emissions. The infra-red emission had
in fact faded away completely by the time the merger took place; thus, demonstrating the
re-energization aspect of these collisions/internal shocks.

The lightcurves shown in figures \ref{fig:doubleloss_lc_rad} and
\ref{fig:doubleloss_lc_ir} illustrate the
lag between the high and the low frequency peaks, already seen from
individual shells. In this case, however, the lightcurves show
the total emission from the entire jet (only two shells for this
scenario; multiple shells case is presented below). 
The radio emission not only lags behind the infra-red emission, but also has a
much lower peak flux value (due to energy losses). Also, the radio
emission rise and decay times are much longer than the infra-red
times. In the case of infra-red, there is a sharp rise in the
emission at the point of shell merger. As mentioned earlier, the
infra-red fades away almost completely by the time the shell collision takes
place. The merger, however, has a compression effect, thus increasing the energy
density of the newly formed shell, causing it to start emitting in
infra-red. The internal energy generated at the collision is
still not sufficient to make the shell optically thick to infra-red; therefore,
we do not see a slow rise in the infra-red flux from the merged
shell. The picture is slightly different for the radio: the merged
shell has high enough energy density that it becomes optically thick
to the radio frequencies, leading to a slow rise in the flux. The other
point to note, for the radio, is that after the second shell is
injected into the jet, it takes a long time before the maximum flux is
reached. This is due to the second shell taking a long time to become
optically thin to the radio frequencies.

\subsection{Multiple ejections}
The multiple ejections set-up is also demonstrated with and without
adiabatic energy losses taking place. When the energy losses are
absent, the internal shocks/collision are also omitted as no energy
replenishment is required; the collisions, however, are modelled when
energy losses are incorporated. The shell properties in the
case of collisions set-up are sampled from a \textit{pseudo}-random
distribution (as described in section \ref{subsec:modelparams}).

\subsubsection{Without energy losses}
If the time gap between ejection is small then an almost continuous
jet can be approximated by multiple ejections. All the shells are
injected with the same properties (time gap, velocity etc.), hence no
collisions take place. The simulation parameters are given in table
\ref{tab:multicolNoloss}. 

\begin{figure}
\includegraphics[width=0.48\textwidth]{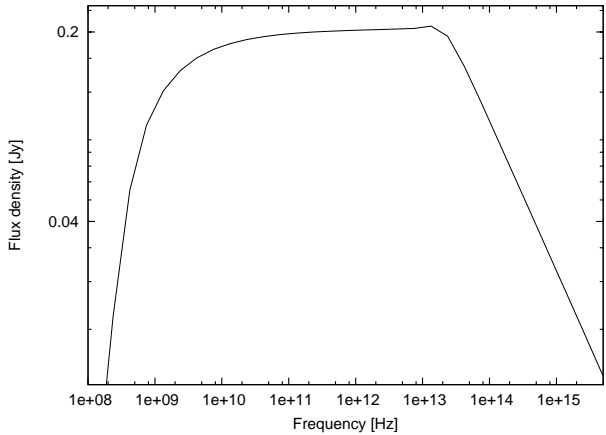}
\caption{Spectrum of a jet modelled using multiple ejections (without
  energy losses). An
  almost flat spectrum is achieved over a large frequency range. It
  should be noted that the low frequency turn over point is dictated
  by the duration of the simulation: longer simulation means a longer
  jet which means a lower frequency turnover. (see table
  \ref{tab:multicolNoloss} for parameters).}
\label{fig:multicol_nuF}
\end{figure} 

\begin{table}
\caption{The parameters used to demonstrate multiple ejections
  (without energy losses).}
\begin{center}
\begin{tabular}{@{\extracolsep{\fill}}lr@{}}
\hline
Parameter & Figures \ref{fig:multicol_nuF}, \ref{fig:multicol_ir},
\ref{fig:multicol_rad}, \ref{fig:multicol_mult}\\
\hline
$\varphi$ & 5$^o$ \\
$\theta$  & 40$^o$ \\
$D$ & $2$ kpc \\
$L_W$ & $1\times 10^{30}$ J/s \\
$\Gamma$ & 2.0 \\
$l_{scale}$ & 0.2 \\
$u_e$ & 0.5 \\
$u_B$ & 0.5 \\
$u_{th}$ & 0.0 \\
$x_{shock}$ & 0.0 ls \\
$E_{frac}$ & 0.01 \\
$\textrm{d}t_{inj}$ & $\sim$1 s \\
Sim. Duration & $5\times10^4$ s \\
\hline
\end{tabular}
\end{center}
\label{tab:multicolNoloss}
\end{table}

\begin{figure}
\includegraphics[width=0.48\textwidth]{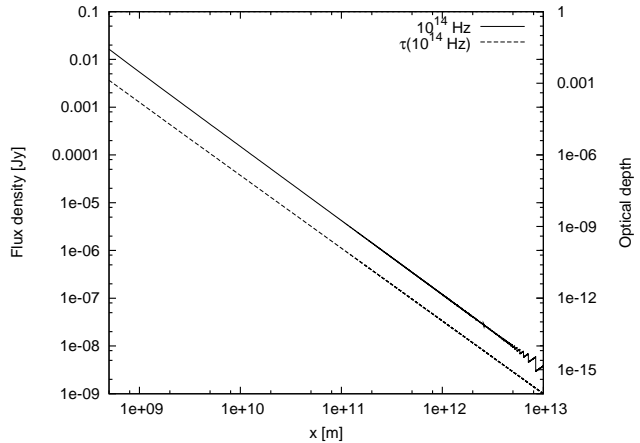}
\caption{The infra-red emission of a multiple ejections jet (without
  energy losses). Optical depth
  corresponding to the infra-red frequency is also plotted (long
  dashed line). The wriggles at the end of the IR spectrum are a
  numerical artefact.(see table
  \ref{tab:multicolNoloss} for parameters).}
\label{fig:multicol_ir}
\end{figure} 

\begin{figure}
\includegraphics[width=0.48\textwidth]{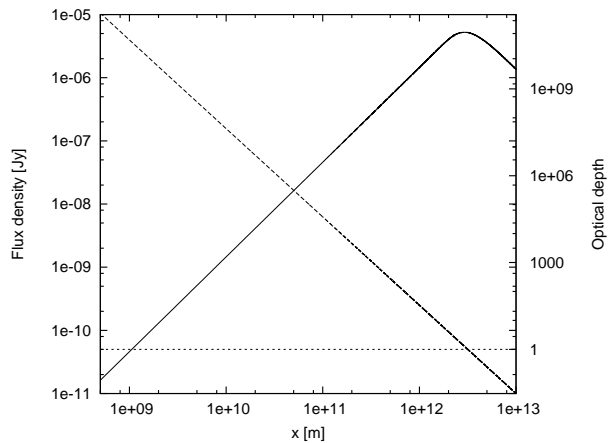}
\caption{The radio emission of a multiple ejections jet (without
  energy losses). Optical depth
  corresponding to the radio frequency is also plotted (long dashed
  line); short dashed line shows where the optical depth $\sim1$.(see table
  \ref{tab:multicolNoloss} for parameters).}
\label{fig:multicol_rad}
\end{figure} 

\begin{figure}
\includegraphics[width=0.48\textwidth]{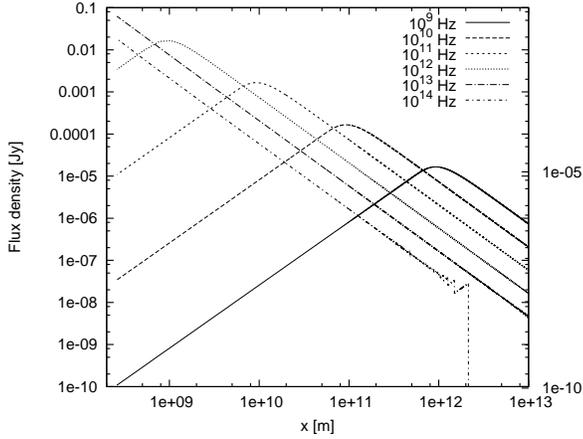}
\caption{The emission as a function of the jet radius for a range of
  frequencies in a multiple ejections jet (without energy losses). (see table
  \ref{tab:multicolNoloss} for parameters).}
\label{fig:multicol_mult}
\end{figure}

The spectrum shown in figure \ref{fig:multicol_nuF} shows how a flat
spectrum (for a specific frequency range) can be recovered when no
energy losses are considered. This set up can be compared to a situation
where one assumes a constant replenishment of any energy losses by an
unknown mechanism \citep{blandford79}. The evolution of a radio
and an IR frequency along the jet can be seen in figures
\ref{fig:multicol_ir} and \ref{fig:multicol_rad} respectively. The two
frequencies show a very different behaviour: IR spectrum 
shows a constant decline while the radio spectrum peaks much further
down the jet. A look at the optical depths for
the two frequencies shows that the injected shells, and ultimately the
jet, is optically thin to infra-red ($\tau_{\nu} << 1$); in the case of the
radio frequency the jet becomes optically thin at a large distance
from the source. (The
same behaviour was observed in the case of single ejections as
well). The emission for a range of
frequencies
(radio $< \nu <$ infra-red) are shown in figures
\ref{fig:multicol_mult}. The different emissions
shown in figure \ref{fig:multicol_mult} follow the
$R_{\tau\approx 1}\propto \nu^{-1}$ relation (as predicted
analytically by \citet{blandford79}).

\subsubsection{With adiabatic energy losses}
The internal shocks are a
possible way to address the problem of replenishing the energy losses
in a jet. In the simulations presented below the shells are expanding
both longitudinally and laterally; therefore the adiabatic losses can
be extremely fast, making the flat spectrum difficult to obtain. The
spectra from simulations where shells are not injected with any
internal energy can be compared with the spectra from the simulation
where shells are injected with internal energy: internal shocks are
the only source of the internal energy production in the former case,
whereas in the latter they serve to replenish the energy losses only.

\begin{figure}
\includegraphics[width=0.48\textwidth]{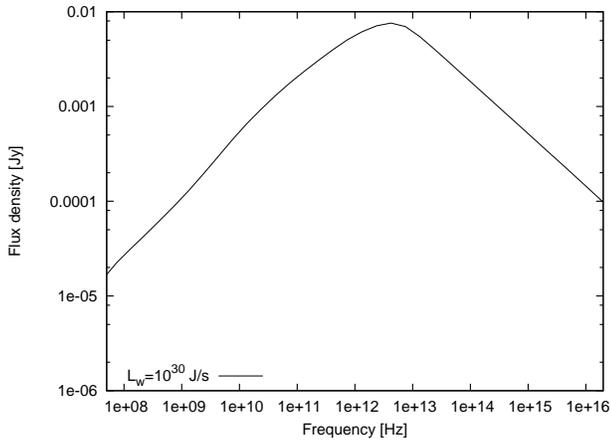}
\caption{Time averaged spectrum ($\sim1k s$) from a multiple ejection jet
  with adiabatic energy
  losses. The shells involved undergo
  lateral and longitudinal expansion. The shells are not injected with
  any internal energy. (see table
  \ref{tab:paramsmultiloss} for parameters).}
\label{fig:multicol_nuFloss}
\end{figure} 

The spectrum shown in figure \ref{fig:multicol_nuFloss} shows a highly
inverted spectrum, when the shells are not injected with any internal
energy. The internal shocks taking place are not sufficient to
produce the internal energy in addition to replenish the energy losses
that are taking place, resulting in an inverted spectrum.

\begin{figure}
\includegraphics[width=0.48\textwidth]{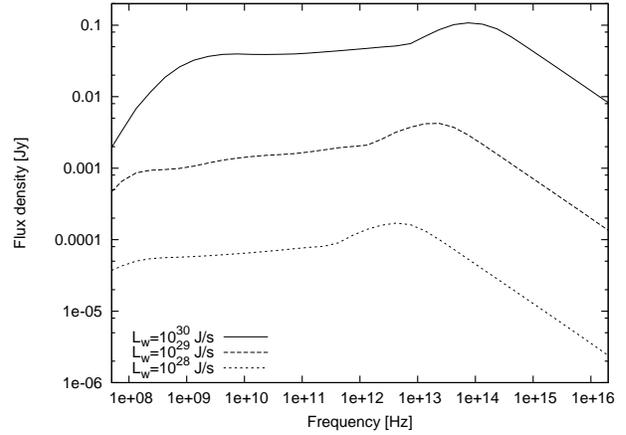}
\caption{Time averaged inverted/flat spectra ($\sim2k s$) from multiple
  ejection jets with the adiabatic energy
  losses. The three spectra correspond to different jet kinetic
  luminosities. The shells involved undergo
  lateral and longitudinal expansion. The shells are injected with
  internal energy, creating a much different spectrum from seen in
  figure \ref{fig:multicol_nuFloss}.(see table
  \ref{tab:paramsmultiloss} for parameters).}
\label{fig:multicol_nuFFlat}
\end{figure}

\begin{figure}
\includegraphics[width=0.48\textwidth]{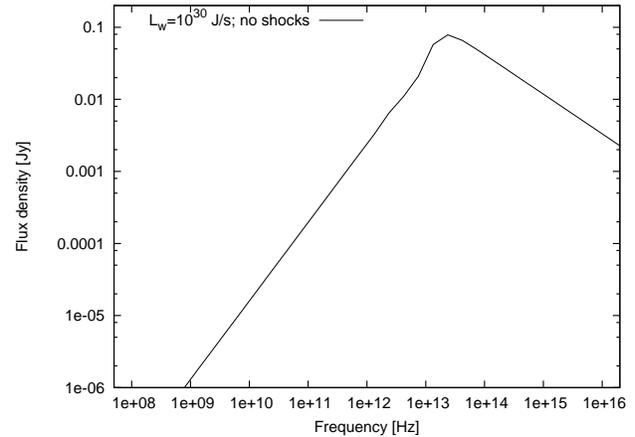}
\caption{Time averaged inverted spectrum ($\sim2k s$) from multiple
  ejection jets with the adiabatic energy
  losses. The adiabatic energy losses are due to the lateral expansion
  of the shells. See table
  \ref{tab:paramsmultiloss} for parameters.}
\label{fig:multicol_nuFNotFlat}
\end{figure}

\begin{table}
\caption{The parameters used for multiple ejections jet with adiabatic
  energy losses.}
\begin{center}
\begin{tabular}{@{\extracolsep{\fill}}lccr@{}}
\hline
Parameter & fig \ref{fig:multicol_nuFloss} 
     & fig \ref{fig:multicol_nuFFlat} & fig \ref{fig:multicol_nuFNotFlat}\\
\hline
$\varphi$ & 5$^o$ & 5$^o$ & 5$^o$\\
$\theta$  & 40$^o$ & 40$^o$ & 40$^o$\\
$D$ & $2$ kpc & $2$ kpc & $2$ kpc\\
$L_W$ & $1\times10^{30}$ J/s & $1\times10^{28 - 30}$ J/s & $1\times10^{30}$ J/s\\
$\Gamma_{min}$ & 1.5 & 1.5 & 2.0\\
$\Gamma_{max}$ & 2.0 & 2.0 & 2.0\\
$l_{scale}$ & 0.2 & 0.2 & 0.2\\
$u_e$ & 0.33 & 0.33 & 0.5\\
$u_B$ & 0.33 & 0.33 & 0.5\\
$u_{th}$ & 0.33 & 0.33 & 0.0\\
$E_{frac}$ & 0.0 & 0.01 & 0.01\\
$\textrm{d}t_{inj}$ & $\sim$1 s & $\sim$1 s & $\sim$1 s\\
Sim. Duration & $5\times10^4$ s& $5\times10^4$ s & $5\times10^4$ s\\
\hline
\end{tabular}
\end{center}
\label{tab:paramsmultiloss}
\end{table}

On the other hand, the spectra given in figure
\ref{fig:multicol_nuFFlat} illustrate how
it is possible to obtain an inverted/flat spectra even with adiabatic
energy losses taking place. In order to achieve this, it is
necessary to inject the shells
with internal energy. In other words, when the internal shocks are used to
produce the internal energy, plus replenish the adiabatic losses, a
flat spectrum is not obtainable; however, when the internal shocks are
used for energy replenishment only, the flat/inverted spectrum is
recovered. It should be noted that the role played by the injected internal energy
($E_{frac} > 0.0$) is somewhat more
complicated as it also influences the longitudinal expansion rate of the shells, thus
how quickly a shell becomes optically thin to different
frequencies. When a shell is not injected with any internal energy
($E_{frac} = 0.0$), the
longitudinal expansion does not begin until after the first
merger; this means the colliding shells are quite small,
causing higher frequencies to dominate the spectrum. Figure
\ref{fig:multicol_nuFNotFlat} shows a highly inverted spectrum for a
jet where the
shells are
injected with some internal energy, but no shocks/mergers are
taking place; the adiabatic energy losses are also being modelled
(this is an almost identical scenario to the one shown in figure
\ref{fig:multicol_nuF}, except for the additional adiabatic energy
losses here). It is clear that both the internal energy and the
shocks re-acceleration are necessary to achieve a flat/inverted
spectrum whilst taking the adiabatic energy losses into account.

We can also note in figure \ref{fig:multicol_nuFFlat} that the flux is
correlated with the jet kinetic luminosity. This is because $E_{frac}$
is scaled according to the relativistic energy of the shell, which is
related to the mass and the bulk Lorentz factor of the shell; the mass is
dependent on the kinetic luminosity (see equation
\ref{eqn:jetkinetic}), which ultimately means that the higher jet kinetic
luminosity creates shells with higher internal energy, thus producing
greater flux. Higher energy density also means that a shell would be
optically thick to higher frequencies. The effects of jet kinetic
luminosity on the flux (and the spectrum) are degenerate with
$E_{frac}$ parameter: a lower luminosity jet, but with the higher
$E_{frac}$ value can produce similar results. This degeneracy extends
to any parameter that influences the internal energy of the shell at
injection; for instance, the jet opening angle and the shell length ($l_{scale}$)
will also influence the form of spectra obtained. The spectra show in
figure \ref{fig:multicol_nuFFlat} conform approximately to the
relation: $F_{\nu}\propto L_{\textrm{W}}^{\sim1.4}$. This is in
agreement with the relation found analytically by \citet{heinz03}, stating:
$F_{\nu} \propto L_{\textrm{W}}^{\sim1.4}$.

The flat/inverted spectra produced have shown other interesting
correlations: both the high and the low frequency turnover points in
the spectrum correspond to certain jet properties. In the case
of the high frequency break, the higher the jet power (mainly
$L_{\textrm{W}}$, but also $E_{frac}$), the higher the break
frequency. The high frequency
break scales approximately as: $\nu_b \propto L_{\textrm{W}}^{\sim
  0.6}$, which is remarkably close to previously observed and
calculated relation of $\nu_b \propto L_{\textrm{W}}^{\sim
  0.7}$ \citep{falcke95, markoff03, heinz03}. The low frequency turnover, on the other
hand, is also
affected by the jet luminosity, but the re-energization by the internal
shocks appears to play the biggest role: both the number of shells
present in the jet and the collision radii of the shells influence the
low frequency turnover.   

\section{Conclusions}
The results presented in this paper show how it is possible to reproduce a
canonical flat spectrum even when a discretized jet is used.  We have
also shown how flat/inverted spectrum is also reproducible if the
internal shocks are used for energy replenishment. If the internal
shocks are used for the initial electron acceleration, on top of
replenishing the energy
losses, the spectra become highly inverted
($\alpha > 0$).

The multiple internal shocks created by multiple ejection into the
jet volume can provide a considerable amount of energy to the shells. The
results show that even with an essentially random distribution of
shells velocities and injection times, adequate
re-energization is possible: the flat/inverted spectrum is
achieved. This is an important result in furthering our understanding
of the jet physics. We have also seen that the high frequency break in flat/inverted
spectra is correlated with the jet power; the lower frequency turnover
shows dependence on the number of shells present in the jet in
addition to their collision radii. Further investigation is
required to quantify fully the relation that may exist between the jet
properties and various break frequencies. However, the break frequency
correlations seen thus far, are in agreement with the theoretical
prediction as well as the observations. 

The results outlined above also hint at being able to tie the timing
properties (X-ray to Infra-red in the case of X-ray binaries) with the
jet physics. It should also be possible to link the shell
properties (such as time gap between ejections and the bulk Lorentz
factor) with
X-ray timing information for example. Using the X-ray lightcurves to drive
such jets, we can then look at the infra-red lightcurves produced,
we have an additional diagnostics for checking self consistency in the
model. This investigation may also show how a ``single blob''
picture may arise, where the radio rise and decay times (plus the flux) are very
similar to the ones for the infra-red \citep{mirabel98,
  fender00b}. This is clearly not compatible
with the single shell picture presented in this paper. It can then be
assumed that it is not a simple scenario like a single large blob
being responsible for the observed  massive
ejection events. We must therefore delve further to try and
understand why this is currently not reproducible with our model.

It is safe to conclude that the results presented in this paper do not
exclude other re-acceleration models, but the internal shocks model
holds much promise in being able to reproduce the often seen
flat/inverted spectra
in addition to opening up the avenue for studying jet timing properties.  

\section*{Acknowledgments}
The authors are grateful to the anonymous reviewer for many helpful
comments. OJ is grateful to STFC for the Ph.D. funding. OJ would also like to thank
Phil Uttley for the insightful discussion and reading the paper. OJ is also
grateful to Tom Maccarone, Tony Bird, Cl\'ement Cabanac, Piergiorgio
Casella and Georgi Pavloski for many useful discussions.

\appendix

\section{The code}
\label{app:params}
We are in principle open and receptive to others wishing (in
collaboration) to utilize our code to model and test different jet
scenarios. If you are interested, please contact the author. 

The following section outlines all the customizable parameters in our
model. With efficiency and expandability in mind, the model is coded
in C++. Effort has been made to minimize dependencies
and use GNU software only. Once compiled, the code can read in all the parameters from a
simulation parameters file and a shell parameters file; for any
subsequent changes to the parameters, the code does not require
re-compilation.  
\\\\
The customizable parameters for the code are as follows:
\\\\
\textbf{Jet Luminosity}:
Used when shell properties are \textit{pseudo}-random. This determines
the shell mass based on how many shells need to be injected. [J/s]
\\\\
\textbf{BLF\_max}:
Used when shell properties are \textit{pseudo}-random. This sets the
mean of a Gaussian distribution to be sampled from. [$\Gamma_{max}$]
\\\\
\textbf{BLF\_min}:
Used when shell properties are \textit{pseudo}-random. This sets the
mean of a Gaussian distribution to be sampled from. [$\Gamma_{min}$]
\\\\
\textbf{shell\_width\_factor}:
Sets the initial shell size, along the jet axis using the relation
outlined in equation \ref{eqn:shellLength}. [no units]
\\\\
\textbf{jet\_opening\_angle}:
The full opening angle of the jet. [degrees]
\\\\
\textbf{source\_distance}: 
Distance to the source being modelled. [kpc]
\\\\
\textbf{EThermal\_frac}:
The fraction of the shell internal energy to be given to the thermal
energy; causes the longitudinal expansion. [no units]
\\\\
\textbf{EelecKin\_frac}:
The fraction of the shell internal energy to be given to the total
electron kinetic energy; determines the power-law distribution
parameters. [no units]
\\\\
\textbf{EMagnet\_frac}:
The fraction of the shell internal energy to be given to the magnetic
energy; affects the magnetic field strength. [no units]
\\\\
\textbf{powerlaw\_index}: 
The power-law index, $p$, of the electron power-law distribution. [no units]
\\\\
\textbf{e\_gamma\_min}:
$\gamma_{min}$ of the power-law electrons. [$\gamma$]
\\\\
\textbf{e\_gamma\_max}:
$\gamma_{max}$ of the power-law electrons. [$\gamma$]
\\\\
\textbf{nu\_min}:
$\nu_{min}$ for the frequencies being modelled; used when
logarithmically spaced frequency range is used. [Hz]
\\\\
\textbf{nu\_max}:
$\nu_{max}$ for the frequencies being modelled; used when
logarithmically spaced frequency range is used. [Hz]
\\\\
\textbf{nu\_points}:
Determines the number of points for the logarithmic frequency grid. [no units]
\\\\
\textbf{individual\_frequencies}
Switch to turn on logarithmically spaced frequency range; takes min. and
max from above. [take integer values: n=off (then uses two frequencies
  below; y=on]
\\\\
\textbf{nu\_1}
Used if only two frequencies being sampled. [Hz]
\\\\
\textbf{nu\_2}
Used if only two frequencies being sampled. [Hz]
\\\\
\textbf{increase\_time\_resolution}
A switch for increasing sampling, for radiative emission, of the jet at the time interval
given below. [takes integer values: n=off (in this case the jet is
  sampled only at ``events''; y=on]
\\\\
\textbf{step\_resolution}
If the above switch is on, this determines the sampling time
interval. [s]
\\\\
\textbf{total\_run\_duration}
Total run time of the simulation. [s]
\\\\
\textbf{shell\_inj\_duration}
The length of time for the shells injection. [s]
\\\\
\textbf{avg\_ejection\_gap}
Sets the mean of the Gaussian distribution for sampling the time
interval between shell injections. [s]
\\\\
\textbf{use\_shell\_file}
A switch for reading a file with shell properties: injection time,
shell mass, shell Lorentz factor, shell width. [takes integer values:
  n=off; y=on (when on, jet\_luminosity, BLF\_max, shell\_width,
  shell\_inj\_duration, and 
  avg\_ejection\_gap are all deactivated)]
\\\\
\textbf{shell\_file}
Name of the file with shell parameters. [should contain 4 columns with
  the corresponding shell properties:injection time,
shell mass, shell Lorentz factor, shell width.]
\\\\
\textbf{write\_results\_file}
A switch to activate writing every time step to a file. [takes integer
  values: n=off; y=on]
\\\\
\textbf{results\_file}
Name of the file for the above switch.
\\\\
\textbf{final\_time\_step}
A switch to activate writing final time step of the simulation. [takes integer
  values: n=off; y=on]
\\\\
\textbf{lightcurve\_file}
Name of the file for writing the light curve data. [Always written by default]
\\\\
\textbf{in\_vacuum\_expansion}
A switch to deactivate adiabatic losses.
\\\\
\textbf{inj\_int\_energy}
A switch to inject the shells with internal energy, after they have
passed the ``shock location''.
\\\\
\textbf{rel\_mass\_frac}
Scale the amount of internal energy given to the shell by a fraction
of the of the shell's relativistic energy.
\\\\
\textbf{shock\_location}
The shock location. Used when injecting with internal energy. [light seconds]
\\\\
\textbf{slow\_energization}
A switch to activate slow energization; the shells are not energized
instantly, but given the energy over a length of time determined by
the shock crossing time. [takes integer values: n=off, y=on]

\bibliographystyle{mn2e}
\bibliography{bibliography}

\begin{thebibliography}{}

\bibitem[\protect\citeauthoryear{{Blandford} \& {K\"onigl}}{{Blandford} \&
  {K\"onigl}}{1979}]{blandford79}
{Blandford} R.~D.,  {K\"onigl} A.,  1979, The Astrophysical Journal, 232, 34

\bibitem[\protect\citeauthoryear{{Cawthorne}}{{Cawthorne}}{1991}]{cawthorne91}
{Cawthorne} T.~V.,  1991, {Interpretation of parsec scale jets}.
pp 187--+

\bibitem[\protect\citeauthoryear{{Corbel} \& {Fender}}{{Corbel} \&
  {Fender}}{2002}]{corbel02}
{Corbel} S.,  {Fender} R.~P.,  2002, The Astrophysical Journal, 573, L35

\bibitem[\protect\citeauthoryear{{Falcke}}{{Falcke}}{1996}]{falcke96}
{Falcke} H.,  1996, The Astrophysical Journal Letters, 464, L67+

\bibitem[\protect\citeauthoryear{{Falcke} \& {Biermann}}{{Falcke} \&
  {Biermann}}{1995}]{falcke95}
{Falcke} H.,  {Biermann} P.~L.,  1995, Astronomy and Astrophysics, 293, 665

\bibitem[\protect\citeauthoryear{{Falcke} \& {Markoff}}{{Falcke} \&
  {Markoff}}{2000}]{falcke00}
{Falcke} H.,  {Markoff} S.,  2000, Astronomy and Astrophysics, 362, 113

\bibitem[\protect\citeauthoryear{{Fender}}{{Fender}}{2001}]{fender01}
{Fender} R.~P.,  2001, Mon. Not. R. Astron. Soc., 322, 31

\bibitem[\protect\citeauthoryear{{Fender}, {Belloni} \& {Gallo}}{{Fender}
  et~al.}{2004}]{fender04}
{Fender} R.~P.,  {Belloni} T.~M.,    {Gallo} E.,  2004, Mon. Not. R. Astron.
  Soc., 355, 1105

\bibitem[\protect\citeauthoryear{{Fender} \& {Pooley}}{{Fender} \&
  {Pooley}}{2000}]{fender00b}
{Fender} R.~P.,  {Pooley} G.~G.,  2000, Mon. Not. R. Astron. Soc., 318, L1

\bibitem[\protect\citeauthoryear{{Heinz} \& {Begelman}}{{Heinz} \&
  {Begelman}}{2000}]{heinz00}
{Heinz} S.,  {Begelman} M.~C.,  2000, The Astrophysical Journal, 535, 104

\bibitem[\protect\citeauthoryear{{Heinz} \& {Sunyaev}}{{Heinz} \&
  {Sunyaev}}{2003}]{heinz03}
{Heinz} S.,  {Sunyaev} R.~A.,  2003, Mon. Not. R. Astron. Soc., 343, L59

\bibitem[\protect\citeauthoryear{{Hjellming} \& {Johnston}}{{Hjellming} \&
  {Johnston}}{1988}]{hjellming88}
{Hjellming} R.~M.,  {Johnston} K.~J.,  1988, The Astrophysical Journal, 328,
  600

\bibitem[\protect\citeauthoryear{{Kaiser}}{{Kaiser}}{2006}]{kaiser06}
{Kaiser} C.~R.,  2006, Mon. Not. R. Astron. Soc., 367, 1083

\bibitem[\protect\citeauthoryear{{Longair}}{{Longair}}{1994}]{longair94}
{Longair} M.~S.,  1994, {High energy astrophysics. Vol.2: Stars, the galaxy and
  the interstellar medium}.
Cambridge: Cambridge University Press, |c1994, 2nd ed.

\bibitem[\protect\citeauthoryear{{Markoff}, {Falcke} \& {Fender}}{{Markoff}
  et~al.}{2001}]{markoff01}
{Markoff} S.,  {Falcke} H.,    {Fender} R.,  2001, Astronomy and Astrophysics,
  372, L25

\bibitem[\protect\citeauthoryear{{Markoff}, {Nowak}, {Corbel}, {Fender} \&
  {Falcke}}{{Markoff} et~al.}{2003}]{markoff03}
{Markoff} S.,  {Nowak} M.,  {Corbel} S.,  {Fender} R.,    {Falcke} H.,  2003,
  Astronomy and Astrophysics, 397, 645

\bibitem[\protect\citeauthoryear{{Marscher}}{{Marscher}}{1980}]{marscher80}
{Marscher} A.~P.,  1980, The Astrophysical Journal, 235, 386

\bibitem[\protect\citeauthoryear{{Mirabel}, {Dhawan}, {Chaty}, {Rodriguez},
  {Marti}, {Robinson}, {Swank} \& {Geballe}}{{Mirabel}
  et~al.}{1998}]{mirabel98}
{Mirabel} I.~F.,  {Dhawan} V.,  {Chaty} S.,  {Rodriguez} L.~F.,  {Marti} J.,
  {Robinson} C.~R.,  {Swank} J.,    {Geballe} T.,  1998, Astronomy and
  Astrophysics, 330, L9

\bibitem[\protect\citeauthoryear{{Panaitescu} \&
  {M{\'e}sz{\'a}ros}}{{Panaitescu} \& {M{\'e}sz{\'a}ros}}{1999}]{panaitescu99}
{Panaitescu} A.,  {M{\'e}sz{\'a}ros} P.,  1999, The Astrophysical Journal, 526,
  707

\bibitem[\protect\citeauthoryear{{Pe'er} \& {Casella}}{{Pe'er} \&
  {Casella}}{2009}]{pe'er09}
{Pe'er} A.,  {Casella} P.,  2009, ArXiv e-prints 0902.2892

\bibitem[\protect\citeauthoryear{{Spada}, {Ghisellini}, {Lazzati} \&
  {Celotti}}{{Spada} et~al.}{2001}]{spada01}
{Spada} M.,  {Ghisellini} G.,  {Lazzati} D.,    {Celotti} A.,  2001, Mon. Not.
  R. Astron. Soc., 325, 1559

\end{thebibliography}

\label{lastpage}

\end{document}